\documentclass{article}    

\usepackage{graphicx}

\title{\textbf{Conscious Pulse II: \mbox{The rules of engagement}}}  
\author{Richard Mould\footnote{Department of Physics and Astronomy, State University of New York, Stony Brook,
\mbox{New York} 11794-3800; http://nuclear.physics.sunysb.edu/ \~{}mould}}  
\date{}    
   
\begin{document}             

\maketitle              

\begin{abstract}

      	This is the final paper in a series that considers the rules of engagement between conscious states and
physiological states.  In this paper, we imagine that an endogenous quantum mechanical superposition is created by a
classical stimulus, and that this leads to a `physiological pulse' of states that are in superposition with one
another.  This pulse is correlated with a `conscious pulse' of the kind discussed in a previous paper (Conscious Pulse
I).  We then add a rule (5) to the four rules previously given.  This rule addresses the effect of `pain'
consciousness on both of these pulses, and in doing so, it validates the ``Parallel Principle" applied to pain.    

\end{abstract}

\section*{Introduction}

 In previous papers, I consider a quantum mechanical superposition of apparatus states in the laboratory system,
where an external observer is present during the time it is being produced \cite{RM1},\cite{RM2},\cite{RM3}.  This
results in a superposition of observer brain states that are correlated with the apparatus.  The laboratory
superposition of macroscopic apparatus states therefore become entangled with an \emph{endogenous superposition} of
macroscopic brain states.  Locally, the components of both the laboratory and the endogenous superpositions are
incoherent in a sense explained below and in ref.\ 1.  It is an effect of environmental decoherence \cite{JZ},
\cite{DG}. 

Four \emph{rules of engagement} are proposed in refs.\ 1-3 that describe how brain states arise from and are related to
apparatus states.  In order to formalize this relationship, it is necessary to distinguish between a conscious brain
state and a \emph{ready} brain state.  The latter is physiologically capable of consciousness but is not conscious; and
furthermore, it cannot become conscious until it is chosen by the stochastic process peculiar to quantum mechanics. 
In ref.\ 3, it is found that a conscious state will quickly become a pulse of closely grouped neighborhood states in
the endogenous superposition.  This is called a \emph{conscious pulse}.  A ready brain state may also appear as a
\emph{ready pulse}.  The pulse of a conscious brain state $\underline{B}$ is written \{$\underline{B}$\}, where the
underline means that it is conscious.  The pulse of a ready brain state $B$ is written $\{B\}$.

In the present paper, there is no laboratory superposition that serves as an external stimulus to the observer. 
Instead, we suppose that the observer is `classically' stimulated, but that that is sufficient to initiate an
endogenous quantum mechanical superposition.  This is possible if there are \emph{seed particles} (molecules or atoms)
within the body that, by virtue of their size and Heisenberg uncertainty, become miniature superpositions that
mushroom into larger ones.  In order to see how this  works, we focus on one particular kind of external
stimulus and one kind of consciousness.  The problem is otherwise too difficult.  I therefore limit consideration to
the case of ``pain" consciousness.  The reason for this choice will be explained in a later section.

\section*{Seed Particles}

 There are several possible seeds.  Henry Stapp proposed that the calcium ions that initiate the release of
neurotransmitters might serve this purpose \cite{HS}. But since the stimulus that I will be discussing produces
pain consciousness, I focus on seeds that are related more directly to pain and the alleviation of pain.  These are
the endorphin molecules and other peptides that move through the blood stream and cerebrospinal fluids seeking opiate
receptors to which they can become attached.  When these molecules attach to a receptor they induce euphoria and/or
analgesia in the subject.  These are suitable seed molecules because they are small enough that their Heisenberg
uncertainty of position grows significantly in the time that it takes for them to move from their point of origin to
their final destination \cite{RM4}.   

The extent to which a given receptor is stimulated by a single migratory seed molecule is therefore uncertain.  We
assign a quantum  number $u$ to the number of receptors that are stimulated by all the seed molecules in the
system.  Due to a receptor's strong interaction with its environment, incoherence is locally assured between components
of the resulting superposition.  These receptors are part of a much wider superposition that includes the seed molecule
and their fluid environment; but when all the non-receptor variables are integrated out of cross terms, the
variable $u$ will identify receptor components that lack the possibility of mutual interference (see Appendix).

\section*{The Initial Distribution and Pulse Formation}

Let the observer be subjected to a classical pain stimulus $S_p$.  An exact physiological response $R$
cannot be classically determined from $S_p$ because of the quantum mechanical uncertainty in the number of opiate
receptors that are occupied by seed molecules at that moment.  This means that the pain stimulus represented by $S_p$
is correlated with a distribution of responses represented by $R_u$, where $u$ is the quantum number 
of occupied receptors.  This includes all of the receptor combinations
that sum to that number.  Each $u$ is correlated with a ready brain state that is called into being by the
interaction.  We will say that $R_u$ includes the entire ``low level" physiology of the observer that leads into the
``high level" ready brain state $B_u$.  Immediately following the interaction, the state of the system is therefore 
\begin{equation}
\Phi(t \ge t_0) = S_p\{X\} + S'_p\Sigma_uR_u\{B_u\}
\end{equation}
where $\{X\}$ is the unknown state of the observer prior to his interaction with the classical stimulus, and where the
primed component in eq.\ 1 is equal to zero at $t_0$.  In order to simplify matters, we will initially assume that
$\{X\}$ is not conscious.  Instead, we will say that the observer is unconscious and is aroused by the painful
stimulant that links him to a single (i.e., non-pulse) ready brain state $B_u$.  The more general case of an unknown
conscious pulse $\{\underline{X}\}$ is discussed at the end of this section.  

Equation 1 is then 
\begin{displaymath}
\Phi(t \ge t_0) = S_pX + S'_p\Sigma_uR_uB_u
\end{displaymath}
where again, the primed component is equal to zero at $t_0$.   The sum $\Sigma_uR_uB_u$ is the endogenous superposition
that has been produced by the seed molecules \footnote{This expression does not preclude the possibility that there
may be other endogenous superpositions embedded in each $R_n$.  One is enough to make the point of this paper.}.

	According to the rules in ref.\ 1, the system is certain to experience a stochastic hit on one of the ready brain
states at a time $t_{sc}$, and this reduces all other states to zero.   
\begin{equation}
\Phi(t = t_{sc} > t_0) = S'_pR_{sc}\underline{B}_{sc}
\end{equation}
Then according to rule (3a) in ref.\ 3, this single state will dissolve into a conscious pulse, giving
\begin{equation}
\Phi(t > t_{sc}) = S'_pR_{sc}\{\underline{B}_{sc}\}
\end{equation}
The details of brain state dissolution are discussed in ref.\ 3.  It is assumed that when the entanglement
$R_{sc}\underline{B}_{sc}$ in eq.\ 2 dissolves into $R_{sc}\{\underline{B}_{sc}\}$ in \mbox{eq.\ 3}, the physiological
state $R_{sc}$ will split into a superposition in which each component is connected to a component of the higher brain
pulse $\{\underline{B}_{sc}\}$, while joining with the single classical state $S'_p$ at the other end.   I will not
bracket $R_{sc}$ as I do brain pulses.  So as in \mbox{ref.\ 3}, it will be understood that the physiological connection
to a brain pulse is itself pulse-like.

The probability that the state $\underline{B}_{sc}$ in eq.\ 2 will be chosen is found by integrating $Jdt$ over the
 time of the interaction, making use of the requirement in ref.\ 3 that  $\int \!\!d\alpha
\underline{B}_r^*\underline{B}_s = \delta(r - s)$.  For a stochastic choice that ranges over non-continuous brain
states such as the receptor number $u$, this expression is 
 $$\int \!\!d\alpha \underline{B}_r^*\underline{B}_s = \delta_{rs}    \hspace{1cm}   \mbox{or}   \hspace{1cm}    
\int \!\!d\alpha\underline{B}_{sc}^*\underline{B}_{sc} = 1$$ 
In addition, the square modulus $s =  \int \!\!d\alpha(S_pX)^*(S_pX) = S_p^*S_p$ inasmuch as \mbox{ref.\ 3} requires
that  
\mbox{$\int
\!d\alpha X^*X = 1$}, so we have $$\mbox{Prob}^{(sc)} = (1/s)S_p^*S_p R_{sc}^*R_{sc} \!\!\!\int \!\!d\alpha
B_{sc}^*B_{sc} = R_{sc}^*R_{sc}$$ The total probability is then found by summing over all possible stochastic choices
$$\mbox{Prob}^{(total)} = \Sigma _{sc}\mbox{Prob}^{(sc)} = \Sigma_{sc}R_{sc}^*R_{sc} = \Sigma _uR_u^*R_u$$

The complete process is shown in fig.\ 1.  Stage 1 shows the distribution $R_u^*R_u$ as a function of the number of
receptors $u$ that are occupied at the time of the interaction, where $u_0$ locates the central number, and $u_{sc}$ is
the stochastic choice that is made during the rise time of the distribution (i.e., during the rise time of
$S'_p$).  The large number of u-states is represented as a continuum in \mbox{fig.\ 1}.   

Stage 2 shows the reduction of the distribution to just $R_{sc}$ at time $t_{sc}$ as given in  eq.\ 2.  The single
conscious state in eq.\ 2 then dissolves into the conscious pulse given in eq.\ 3, and this causes the connected
physiological state to fan-out into a connecting \emph{physiological pulse} as shown in stage 3 of fig.\ 1. 
Normalization is not preserved in this reduction.  

\vspace{.2cm}
\begin{figure}[h]
\centering
\includegraphics[scale=0.8]{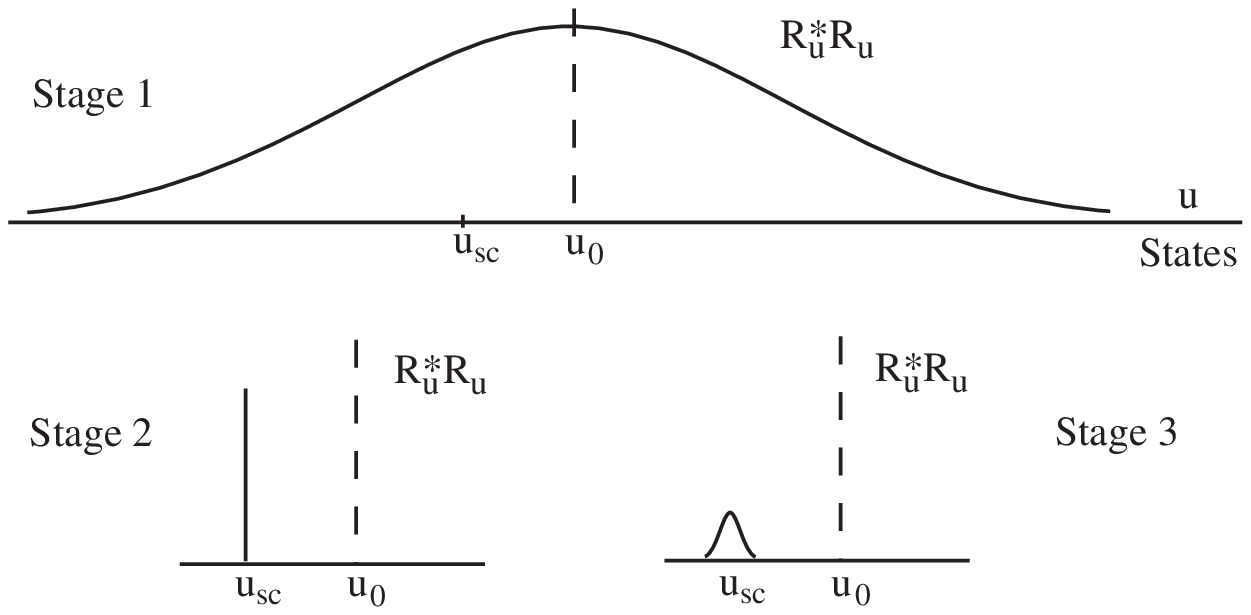}
\center{Figure 1}
\end{figure}

If the unknown pre-interaction state in eq.\ 1 is a conscious observer $\{\underline{X}\}$, then the interaction prior
to $t_{sc}$ will be  
\begin{displaymath}
\Phi(t \ge t_0) = S_p\{\underline{X}\} + S'_p\Sigma_uR_u\{\underline{B}_u\}
\end{displaymath}
where the prime component is again zero at $t_0$.  Substituting the expression for $\{\underline{B}_u\}$ in   eq.\ 2 of
ref.\ 3 gives  
\begin{displaymath}
\Phi(t \ge t_0) = S_p\{\underline{X}\} + S'_p\Sigma_uR_u\int \!\!du'F_u(u')\underline{B}_{u'}
\end{displaymath}
or
\begin{displaymath}
\Phi(t \ge t_0) = S_p\{\underline{X}\} + S'_p\int \!\!du'P_{u'}\underline{B}_{u'}
\end{displaymath}
where $P_{u'} = \Sigma_uR_uF_u(u')$.  A stochastic hit on a value of $u'$ at time $t_{sc}$ will then yield
\begin{equation}
\Phi(t = t_{sc} > t_0) = S'_pP_{sc}\underline{B}_{sc}
\end{equation}
The difference between eq.\ 2 and eq.\ 4 is that the term $R_{sc}$ in eq.\ 2 is replaced by a more a general
physiological expression given by $P_{sc} = \Sigma_uR_uF_u(sc)$.  This uses values of $R$ that are associated with the
conscious distribution, and these may be different from the $R$s associated with the unconscious distribution.  Despite
these differences, stage 2 will always appear as a single state at the stochastically selected site $u_{sc}$ of a
physiological distribution $P_{u'}^*P_{u'}$ similar to the one shown in \mbox{fig.\ 1}; and the final physiological
reduction pulse will always appear like the one shown in stage 3.  The final state will generally take the form
\begin{equation}
\Phi(t \ge t_{sc}) = S'_pP_{sc}\{\underline{B}_{sc}\}
\end{equation}
where the entanglement $P_{sc}\{\underline{B}_{sc}\}$ connects every component of the brain pulse with the
physiological pulse associated with $P_{sc}$.

The probability that the pulse $\{\underline{B}_{sc}\}$ in eq.\ 5 will be chosen is found by integrating $Jdt$ over the
time of the interaction.  In this case, $(sc)$ is assumed to be a continuous variable.  The total probability
is then
$$\mbox{Prob}^{(total)} = (1/s)S_p^*S_p \int \!\!d(sc) P_{sc}^*P_{sc} \int \!\!d\alpha
\{\underline{B}_{sc}\}^*\{\underline{B}_{sc}\} = \int \!\!d(sc)P_{sc}^*P_{sc}$$
where we use  $\int \!\!d\alpha \{\underline{B}_{sc}\}^*\{\underline{B}_{sc}\} = 1$ from ref.\ 3.

In general for an endogenous superposition, we conclude that the stochastic choice of a ready brain state and its
dissolution into a conscious pulse as per rules (3) and (3a) will result in a physiological pulse (in fig.\ 5) that is
correlated with the chosen conscious brain pulse.

\section*{The Parallel Principle and Pain Consciousness}

	The reason I have chosen to focus on pain consciousness is related to my belief in the validity
of the
\emph{parallel principle}, and the fact that pain provides an excellent example of how that principle might work.   

	It is generally accepted that the subjective world of our personal experience corresponds in critical ways with the
objective world that exists outside of ourselves.  It is assumed that formal relationships can be found in the
subjective world that parallel formal relationships that exist in the objective \mbox{world - and} this is the basis of
epistemology in physics.  Von Neumann calls it the \emph{\mbox{psycho-physical} parallelism}.   Why a psycho-physical
parallelism should exist at all is an open question, inasmuch as these two separate realms of reality have such
different natures.  There is no obvious reason why one of these worlds should pay attention to the fortunes or
machinations of the other.  Responding to this point, Leibniz claimed that one is compelled to believe in a
\mbox{Pre-established} Harmony between the two worlds that is arranged by God.  

	Opposed to this, the parallel principle says that the psycho-physical parallelism is the consequence of natural
evolutionary processes.  It is claimed here that the conscious evolution of a species develops in parallel with the
physiological evolution of the species; and for this to happen, there must be an \emph{interaction} between
consciousness and physiology.  The general mechanism for this interaction is described in a previous paper
\cite{RM5}.

	The rules that have been considered so far allow one to believe that consciousness is only epiphenomenal; that is,
that it may be regarded as an insubstantial by-product of a   rule (3) reduction that has no influence of its own. 
But for the parallel principle to work, consciousness must be influential.   It must do something.  Its presence must
give an evolutionary preference to one kind of physiology over another, and to achieve this we adopt rule (5).  

\vspace{0.5 cm}
\noindent
\textbf{Rule (5)}\emph{ If two states within a conscious pulse represent different degrees of pain, then the square modulus of the
state with lesser pain will increase at the expense of the state with greater pain.}
  
\vspace{0.5 cm}

	Let the pulse on the left in fig.\ 2 be the conscious brain pulse in eq.\ 5.  When this pulse is formed soon after
$t_{sc}$, it is centered over $u_{sc}'$.  But the effect of rule (5) is to subsequently move the pulse to the
right as indicated by the arrow in that figure.  

\begin{figure}[h]
\centering
\includegraphics[scale=0.9]{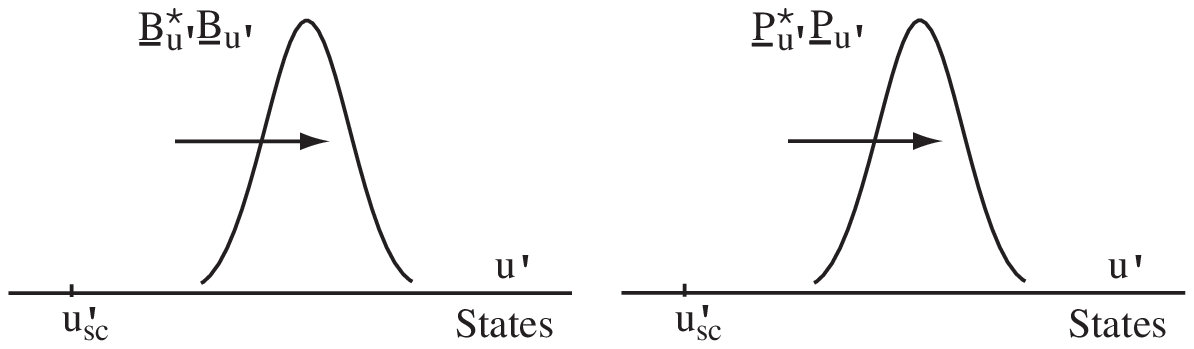}
\center{Figure 2}
\end{figure}

 Recall that $u$ (now $u'$) represents an increasing numbers of opiate receptors that are occupied by seed
molecules.  Therefore, states on the right-hand edge of the conscious pulse have a greater population of occupied
receptors than the states on the left-hand edge.  This means that the right edge states are less painful than the left
edge states.  Rule (5) then requires that right edge states grow in magnitude at the expense of left edge states, so
current flows from left to right.  The net effect is to shift the pulse to the right as shown.   

	The brain pulse $\{\underline{B}\}$ in eq.\ 5 is correlated with the physiological pulse appearing in that equation,
so when the former moves, the latter moves with it.  This is shown in the pulse on the right in fig.\ 2.  It is
also centered over $u_{sc}'$ at $t_{sc}$, and it also moves to the right because of its connection with the
conscious brain pulse. 

	I claim that the mechanisms of evolution will work together with rule (5) to promote the creation of a
psycho-physical parallelism applied to pain.  The rightward drift of the physiological pulse in fig.\ 2 will have
behavioral consequences that may or may not benefit the species.  Remember that each physiological state $P_{u'}$
includes a response to $K_{u'}$ of the entire organism leading up to the high level brain state $B_{u'}$.  The
conscious pulse drift dictated by rule (5) therefore results in a behavioral drift.  The resulting behavioral change
might be trivial or it might be significant; and nothing has been said that would give us a clue as to what that
change might be.  But we can say that if the induced behavior is harmful, then the species will become extinct.  If it
is beneficial, then the species will survive with an instinctive physiological response to $K_{u'}$ that is associated
with a psychological avoidance of pain.  Rule (5) therefore provides a platform from which a psycho-physical
parallelism can be launched.  It establishes a mechanism that makes the parallel principle possible.  Further details
are found in ref.\ 8.

\section*{Causal Influence and Equilibrium}

		The physiological pulse is a superposition whose leading edge components grow at the expense of its trailing edge
components; so as it moves to the right, more and more receptors will be occupied by seed molecules.  This
physiological movement is not caused by physiological terms that are present in the Hamiltonian, except those that
keep the connection between the physiological states $P$ and the upper brain states $B$.  The causal `push' behind this
movement comes from the extra-physical influence of consciousness itself, enforced by rule (5).  It is important to
note that consciousness cannot be thought of as something that is equivalent to just ``another" physiological package. 
It is not a euphemism for an extended physiological mechanism that renders it epiphenomenal after all.  If that were
so, then there would be no reason why consciousness should be shaped by evolution.  If the substance of consciousness
makes no difference to the objective world, then it will make no difference if it does or does not mirror the
objective world.  Therefore, rule (5) must refer directly to the properties of consciousness (in this case the `pain'
in pain consciousness) in order for the parallel principle to work.  

While consciousness is said to override the Hamiltonian as described in rule (5), it is only a partial influence,
competing with the more familiar physiological influences.  So the pulses in fig.\ 2 will not continue indefinitely
to the right.  I assume that these two pulses will always remain correlated, but that opposing tendencies inherent in
the Hamiltonian will eventually bring them to a halt.  For instance, increasing values of $u$ means that there are
greater numbers of occupied opiate receptors.  But the initial distribution in stage I of fig.\ 1 shows that there are
a limited number of seed particles that are available for that purpose.   So the pulse cannot move too far to the
right.  There will be a final equilibrium between the influence introduced by rule (5) and all the other physiological
influences contained in the Hamiltonian. 

	Because rule (5) establishes a causal influence of pain consciousness on physiology, it should be possible in
principle to test that rule experimentally.  I have previously suggested two experiments that purport to test the
existence of the above `pulse drift' among pain states \cite{RM6}.  The experiments use either a PET scan with human
subjects experiencing pain, or autoradiography with rats experiencing pain.  If one of these tests proves to be
positive, I believe that will confirm the account given in this paper as to how a primitive a psycho-physical
parallelism is established.  That will also support my claim that all five of the rules given in these papers are
correct ``rules of engagement" between conscious brain states and physiology.

\section*{A Previous Experiment Is Explained}

			In 1999, the author performed an experiment in which a $\beta$-source was used to create a two-component
superposition, one of which gave the author a painful electric shock, and the other of which gave no shock (see ref.
7).  The idea was to see if the pain consciousness induced by one component of this externally created superposition
would be instrumental in suppressing that component.  At the time, it seemed possible that the effect of pain
consciousness on an external superposition might be directly observable in this way, but the result of the experiment
was negative.  This suggested that that effect of pain is felt at a deeper level, although it was not clear at the
time why that should be so.  The present model shows why that is so.

Painful states are designated by the variable $u$ (left side of fig. 3), and the non-painful states are designated by
$u''$(right side of fig.\ 3).  If the $\beta$-source chooses the painful stimulus with a frequency that is equal to
the non-painful stimulus, then according to my results, the probability that a painfully conscious pulse will arise on
the left in fig.\ 3 will be equal to the probability that a non-painfully conscious pulse will arise on the
right\footnote{The variable $n''$ is not physically interpreted, so its distribution in fig.\ 3 need not look like that
of $n$.  It does not even have to be a quantum superposition.  It is important only that non-painful states are equally
probable with painful states.}.  It is only \emph{after} the stochastic decision has taken place that the differential
influence of pain consciousness can cause a physiological effect.  If a conscious pulse arises among the painful
states on the left side of fig.\ 3, it will drift to the right.  If the pulse develops on the right side of the
figure among the non-painful states, it will not drift in either direction because those states are assumed to be
neutral on a pleasure/pain scale.

\begin{figure}[h]
\centering
\includegraphics[scale=0.85]{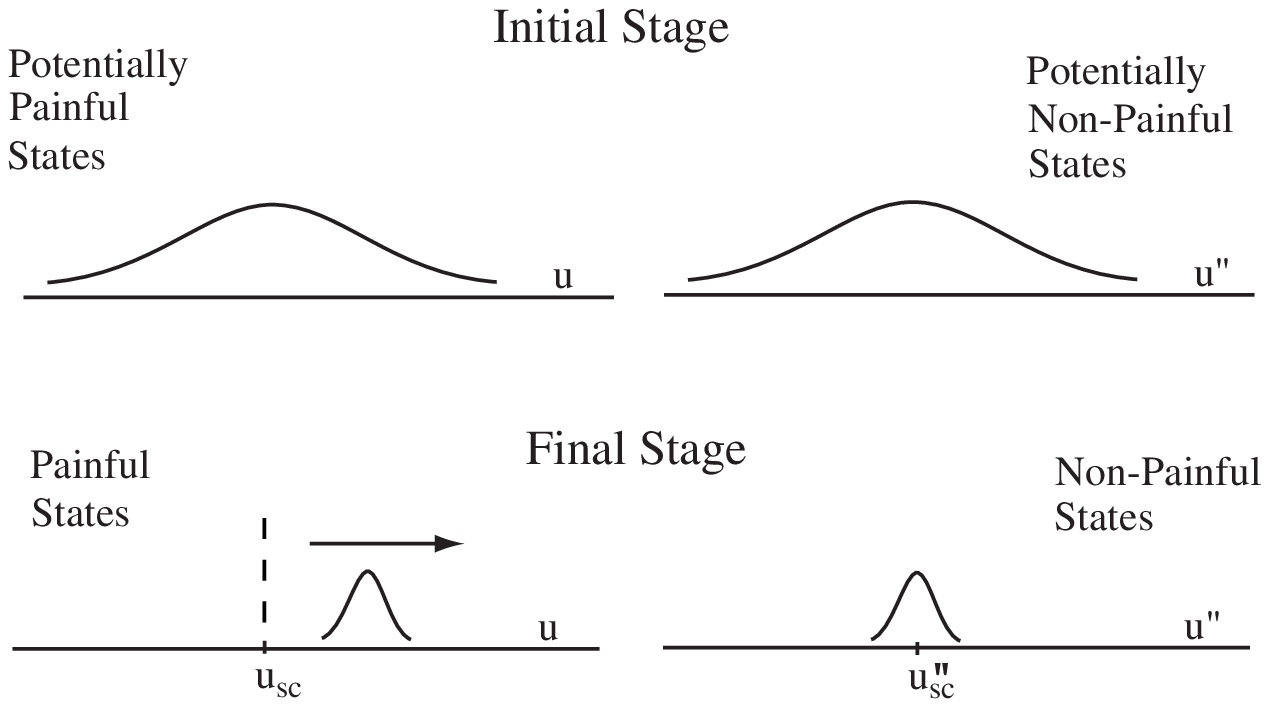}
\center{Figure 3}
\end{figure}

Apparently the differential influence of pleasure or pain consciousness can only occur after the stochastic
choice has located the initial position of the conscious pulse.  The deeper level is then to be found within the
post-stochastic conscious pulse.  This result is consistent with my 1999 experiment; although of course, this
experiment did not allow the drift to be observed.

\section*{Appendix}

				We will say that the classical volume $V_0$ of a molecule is the smallest volume that it can have, consistent with
its intrinsic structure.  If left by itself, its wave function will expand \emph{via} Schr\"{o}dinger because of its
momentum uncertainty, reaching a volume $V$ in time $T$.  The moleculeÕs position within the volume $V$ will be
entirely uncertain, \emph{and} the wave function covering the volume will be \emph{spatially coherent}.  That is, any
spatially identifiable part of the wave is related to any other spatially identifiable part by a well-defined phase
relationship.  

If the molecule is immersed in a stationary liquid, its expansion will be severely limited by the environmental forces
that constrain its motion, thereby limiting the volume that it can reach in time $T$.  However, if the liquid is
undergoing turbulent or laminar flow, then the molecule's volume might very well reach a volume $V$ in time $T$, except
that it will be broken up and widely distributed.  As before, the position of the molecule will be entirely uncertain
within that volume; but in this case, the wave function will be spatially \emph{incoherent}.  That is, when the
environment variables are integrated out of the cross terms between different (spatial) parts of this single molecule,
the result will be zero - indicating incoherence between these parts\footnote{This is generally expressed by saying
(ref. 4) that when the environmental variables are traced out of the  density matrix of the total system, the remaining 
density matrix of the  molecular subsystem is an `improper' mixture.}.  This means that the wave function will be a
\emph{locally incoherent superposition} in the sense that it will not display interference between the molecule's
spatially separated parts (beyond its intrinsic size), even if those parts are adjacent to one another.  The reason for
this break-up of the molecule's wave function is the thermal pounding given to it by its liquid environment.  This
incessant interaction between the liquid and the molecule therefore leads to an environmental decoherence that
disassociates different parts of the wave function from one another.      

The seed molecules we are considering have a maximum atomic mass of 10,000 u and a classical width of at most 10 nm. 
Therefore, their minimal quantum mechanical uncertainty in velocity in one direction will be \mbox{$\Delta v$ = 0.6
mm/sec}, assuming that we begin with a molecule of classical size.  These molecules are carried along by blood and/or
cerebrospinal fluid in turbulent or laminar flow, so their displacement $L$ along that line of flow is likely to be as
much as \mbox{$\Delta v\Delta t = 60 \mu$m} in just 0.1 s.  Therefore, the wave function of this molecule will be an
incoherent quantum mechanical superposition of classically sized molecules that are spread out over this macroscopic
distance.  The particle's position will be uncertain to that extent.

Since a single seed molecule can spread itself out over a large number of opiate receptors in this way, it is able to
produce an uncertainty of stimulation in a widely disbursed group of receptors.  This means that a single molecule can
give rise to an endogenous superposition of receptors like the one in fig.\ 1, and that that superposition will be
locally incoherent when its environmental influences (including the seed molecules) have been integrated out of any
cross terms.

\end{document}